\documentclass{article}
\usepackage{spconf,amsmath,graphicx}
\usepackage[textfont=it,tableposition=top]{caption}
\usepackage{amssymb, amsmath,bm}
\usepackage{textcomp}
\usepackage{booktabs}
\usepackage{siunitx}
\usepackage{xspace}
\usepackage{url}
\usepackage{tipa}
\usepackage{enumitem}
\usepackage{subcaption}
\usepackage{booktabs} 
\usepackage{siunitx}  
\usepackage{multirow} 

\usepackage{pgfplots}
\pgfplotsset{width=6cm, compat=1.4}
\usepackage{multirow}

\usepackage{pifont}

\newcommand{\norm}[1]{\left\lVert#1\right\rVert}

\title{noise robust distillation of self-supervised
speech models via correlation metrics}
\name{\begin{tabular}{c} 
Fabian Ritter-Gutierrez$^{1,2}$, Kuan-Po Huang$^{3,4}$, Dianwen Ng$^{1,5}$, Jeremy H.M. Wong$^2$, Hung-yi Lee$^3$, \\ Eng Siong Chng$^1$, Nancy F. Chen$^2$
\end{tabular}}
\address{
  $^1$Nanyang Technological University    
  $^2$Institute for Infocomm Research (I2R)
  $^3$National Taiwan University\\    
  $^4$ASUS Intelligent Cloud Services
  $^5$Speech Lab of DAMO Academy, Alibaba Group
}
\begin{document}
\ninept
\maketitle

\begin{abstract}
Compared to large speech foundation models, small distilled models exhibit degraded noise robustness. The student's robustness can be improved by introducing noise at the inputs during pre-training. Despite this, using the standard distillation loss still yields a student with degraded performance. Thus, this paper proposes improving student robustness via distillation with correlation metrics. Teacher behavior is learned by maximizing the teacher and student cross-correlation matrix between their representations towards identity. Noise robustness is encouraged via the student's self-correlation minimization. The proposed method is agnostic of the teacher model and consistently outperforms the previous approach. This work also proposes an heuristic to weigh the importance of the two correlation terms automatically. Experiments show consistently better clean and noise generalization on Intent Classification, Keyword Spotting, and Automatic Speech Recognition tasks on SUPERB Challenge.

\end{abstract}
\begin{keywords}
 Self-supervised learning, robustness, correlation, speech recognition, SUPERB.
\end{keywords}

\section{Introduction}

Self-supervised learning (SSL) speech models (a.k.a Speech Foundation Models), which typically have a large number of trainable parameters, have shown promising performance in various downstream speech tasks \cite{SSLreview}, including Intent Classification (IC), Keyword Spotting (KS), and Automatic Speech Recognition (ASR) \cite{superb, wavlm, hubert, w2v2, w2v2light,data2vec2}. However, deploying such models on resource-constrained devices faces practical challenges due to limitations in memory, both in terms of storage and computation. To address this challenge, knowledge distillation (KD) techniques \cite{jinyu_distill,hintonDistill, distilhubert, lighthubert} have been employed to create compressed versions known as lightweight student models. In KD, the student learns from the representations of the original large self-supervised teacher model. Nonetheless, it has been observed in \cite{RobustDistilHuBERT} that these distilled student models may experience performance degradation when the speech signal is corrupted by different types of noise, such as reverberation and background noise, limiting their practical use in noisy environments.

To tackle this challenge, \cite{RobustDistilHuBERT} suggests conditioning the pre-training stage with noise, minimizing the discrepancy between the teacher and student models. A speech input augmented with different noise distortions is provided to both the teacher and student models. Then, the original distillation objective function between the two distorted representations is used to promote noise invariance. Specifically, this method has demonstrated a certain degree of robustness to in-domain training noise. Nevertheless, the evaluation of the noise robustness to out-of-domain noise has not been conducted, leaving it uncertain whether such a method can achieve robust noise invariance with unseen distortions.

In this study, the primary objective is to enhance the generalization capability of the student model, specifically for general noisy speech applications. This involves evaluating the model's performance on out-of-domain noises. The present work proposes to use a correlation based criterion, motivated by Barlow Twins (BT) objective \cite{barlowtwins} into the knowledge distillation framework to improve noise robustness. The original BT objective aims to create representations that are both invariant to distortions and disentangled along the feature dimension. Conventionally, two identical models will receive the same input with different augmentations. In this work, the proposed method involves maximizing the diagonal elements of the cross-correlation matrix between the frozen teacher and trainable student. By computing a cross-correlation matrix of the encoded representations between these networks, we minimize it towards an identity matrix. This optimization learns teacher behavior by driving the diagonal elements of feature correlation to converge to 1. Moreover, it promotes disentangled representations by pushing the off-diagonal elements of the cross-channel dimension feature's correlation towards 0.

However, it is important to note that achieving high cross-correlation on the diagonal elements between teacher and student representations does not necessarily imply that the student is distortion-invariant. It could indicate that the distortions similarly affect both teacher and student models. To address this issue, an additional self-correlation term on the student's representations is proposed to reduce the self-correlations within the student representations. The self-correlation term helps to generate more distortion-invariant representations and enhances disentanglement. 

The motivation for using correlation metrics for noise robust distillation comes from classical Digital Signal Processing (DSP) literature \cite{rabiner79,oppenheim} that uses cross-correlation metrics to compute the similarity between two signals. The intuition is that a correlation metric for distillation pre-training can detect similar patterns between the signal received by the teacher and the student and hence promote robustness on the student representation.

The results obtained demonstrate that this correlation-based method achieves better generalization on downstream tasks such as Intent Classification (IC), Keyword Spotting (KS), and Automatic Speech Recognition (ASR) in both clean and out-of-domain distorted scenarios, as evaluated on the SUPERB benchmark \cite{superb}.
Additionally, when performing parameter optimization of each term of the proposed method, we observed a trade-off between clean and noise generalization. For this reason, we also propose a simple heuristic method to automatically weigh the importance of the off-diagonal minimization of the cross and self-correlation matrix based on the Signal to Noise Ratio (SNR) received by the teacher and the student, respectively. Reported results show even better performance on clean and noise setups under this heuristic approach.

 In contrast to previous works \cite{bt_speaker_rec, bt_er, delores,ng2023hubert} inspired by Barlow Twins, this study deviates in several aspects. Firstly, we do not incorporate a high-dimensional projector network, reducing the additional trainable parameters requirement. Additionally, unlike prior approaches, we retain the time dimension information rather than averaging it prior to cross-correlation computation. Furthermore, this work considers the scenario of knowledge distillation, where the teacher model remains frozen while the student model adopts a significantly smaller network architecture. The distinction in architecture size between the teacher and student models is an important aspect of our approach as well as keeping the pre-training framework simple without additional trainable parameters.

\section{Related work}
While noise robustness in speech applications has been explored \cite{jinyuli2017,Manohar2018ATL,Meng2018AdversarialTL}, not much work has focused on the noise adaptability of a small self-supervised distilled model. Conversely, to improve the noise robustness of an SSL large model, \cite{improvingNoiseContrastiveSpeech} has implemented two losses. The first loss is a standard contrastive loss between the artificial noisy speech and the second loss is a speech reconstruction model between the contextual representations of the SSL model encoder and the clean waveform. This work focuses on the study of noise robustness on the ASR task only, leaving uncertain the transferability of the method to other downstream tasks. Additionally, the work \cite{zhu2022noise} improves noise ASR robustness of a Wav2vec 2.0 \cite{w2v2} model while preserving clean speech ASR performance. The method feeds artificially generated noisy speech to the encoder and uses the clean speech version as the target. Motivated by \cite{improvingNoiseContrastiveSpeech, zhu2022noise,robustw2v2}, the paper \cite{kuanpocontinualtraining} performs continual pre-training on top of domain adversarial training to improve the noise robustness of the HuBERT model \cite{hubert}. The performance has been evaluated on the SUPERB Benchmark \cite{superb}. 

To improve noise robustness in the self-supervised distilled model, \cite{RobustDistilHuBERT} explores the robustness of a DistilHuBERT model by conditioning the pre-training stage with noise. Both teacher and student are fed different noise distortions and standard distillation loss between the representations is applied to promote noise invariance. Evaluations are done under different tasks on the SUPERB Challenge Benchmark. Nevertheless, the work seems to lack analysis under out-of-domain noisy conditions. Finally, \cite{robustdistiller} proposes a similar strategy to \cite{RobustDistilHuBERT} but adds a speech enhancement head aiming at reconstructing the clean speech waveform from the learned representations of the student. While the proposed method in \cite{robustdistiller} shows some improvement in noise robustness, the addition of the speech enhancement head incurs an increase in trainable parameters during pre-training.

Some works have attempted to adapt BT objective to speech/audio domains. In \cite{bt_speaker_rec}, the same BT objective as in the original paper \cite{barlowtwins} is used as an auxiliary loss for a speaker classification task. In \cite{bt_er}, BT objective is also used as an auxiliary loss for an emotion recognition task, and similarly as in \cite{bt_speaker_rec}, the architecture used suits a Computer Vision (CV) task rather than a speech processing one. The works \cite{delores,audiobt} explores the use of BT for audio classification. Both proposals make use of the high-dimensional projector layer and a siamese network architecture. Finally, \cite{ng2023hubert} explores BT objective for noise robustness on HuBERT.

This paper differs from previous approaches in several aspects. First, previous work relies on a flattening operation to feed the audio to a high-dimensional projector layer before computing the cross-correlation matrix. The proposed method in this paper does not, thus avoiding extra trainable parameters during pre-training. Secondly, all previous approaches use the same neural network topology and feed two distorted views of the speech signal to the Siamese architecture before computing the cross-correlation term. In this work, different neural network topologies are used, and the focus is on the study of distillation, where the teacher is not trainable. Finally, differently from \cite{bt_speaker_rec,bt_er,delores,audiobt}, this paper adds a self-correlation term over the trainable student model to further improve noise invariancy and feature dimension decorrelation.

\section{preliminary works}
\subsection{DistilHuBERT}\label{sec:distillation}

Knowledge distillation trains a student model to adopt the behavior of a teacher model \cite{jinyu_distill,hintonDistill}. This work follows DistilHuBERT \cite{distilhubert} to have a direct comparison with \cite{RobustDistilHuBERT} on the effectiveness of the proposed method. DistilHuBERT consists of a sub-network of HuBERT base \cite{hubert}. Namely, let $\mathbf{F}^N$ represent the sub-network, with $N$ denoting the number of transformer layers in the encoder. Here, $\mathbf{F}^N$ differs from HuBERT only in the number of encoder layers. In practice, previous works \cite{distilhubert,RobustDistilHuBERT} have set $N=2$. Let $\mathbf{x}$ represent an input speech utterance. Let  $\mathbf{h}^l \in \mathbb{R}^{T \times D}$ represent the $l$-th hidden layer of the teacher, with $T$, the number of frames and $D$, the feature dimension. DistilHuBERT aims to predict the $l$-th hidden representation $\mathbf{h}^l$ from the teacher as,

\begin{align}
    \mathbf{z} &= \mathbf{F}^2(\mathbf{x})\\ 
    \hat{\mathbf{h}}^l &= \mathbf{p}^l (\mathbf{z}),
    \label{eq:distill_forward_prop}
\end{align}
where $\mathbf{z} \in \mathbb{R}^{T \times D}$ is the last hidden representation of the student model $\mathbf{F}^2$, $\mathbf{p}^l (\mathbf{z})$ represents the ``$l$-th" prediction head over the ``$l$-th" hidden layer $\mathbf{h}^l$.
This work uses the same prediction heads as in \cite{distilhubert,RobustDistilHuBERT}, namely
  $\mathbf{p}_4$,$\mathbf{p}_8$,$\mathbf{p}_{12}$ is used.

In the original paper \cite{distilhubert}, $\mathbf{F}^2$ is trained by interpolating L1-loss and cosine similarity between the predicted representations  $\hat{\mathbf{h}}^l$ and the teacher representations $\mathbf{h}^l$, Namely, the KD loss is defined as,
\begin{align}
       \mathcal{L}_{\text{KD}} &=  \sum_{l\in\{4,8,12\}} {\mathcal{L}^l_\text{1}} - \gamma {\mathcal{L}^l_{\text{cos}}}\\
       {\mathcal{L}_\text{1}}^l &=  \sum_{t=1}^T \frac{1}{D}\norm{\mathbf{h}_t^{l} - \hat{\mathbf{h}}_t^{l}}_1\\
       {\mathcal{L}_{\text{cos}}}^l &=  \sum_{t=1}^T  \log \sigma (\cos(\mathbf{h}_t^{l}, \hat{\mathbf{h}}_t^{l})), 
\end{align}
with $\cos(\cdot, \cdot)$ the cosine similarity function, $\sigma(\cdot)$ the sigmoid function, and $\gamma$ a constant hyperparameter that controls the importance of the cosine similarity term.

\begin{figure*}[!th]
  \centering
  \includegraphics[width=0.95\linewidth]{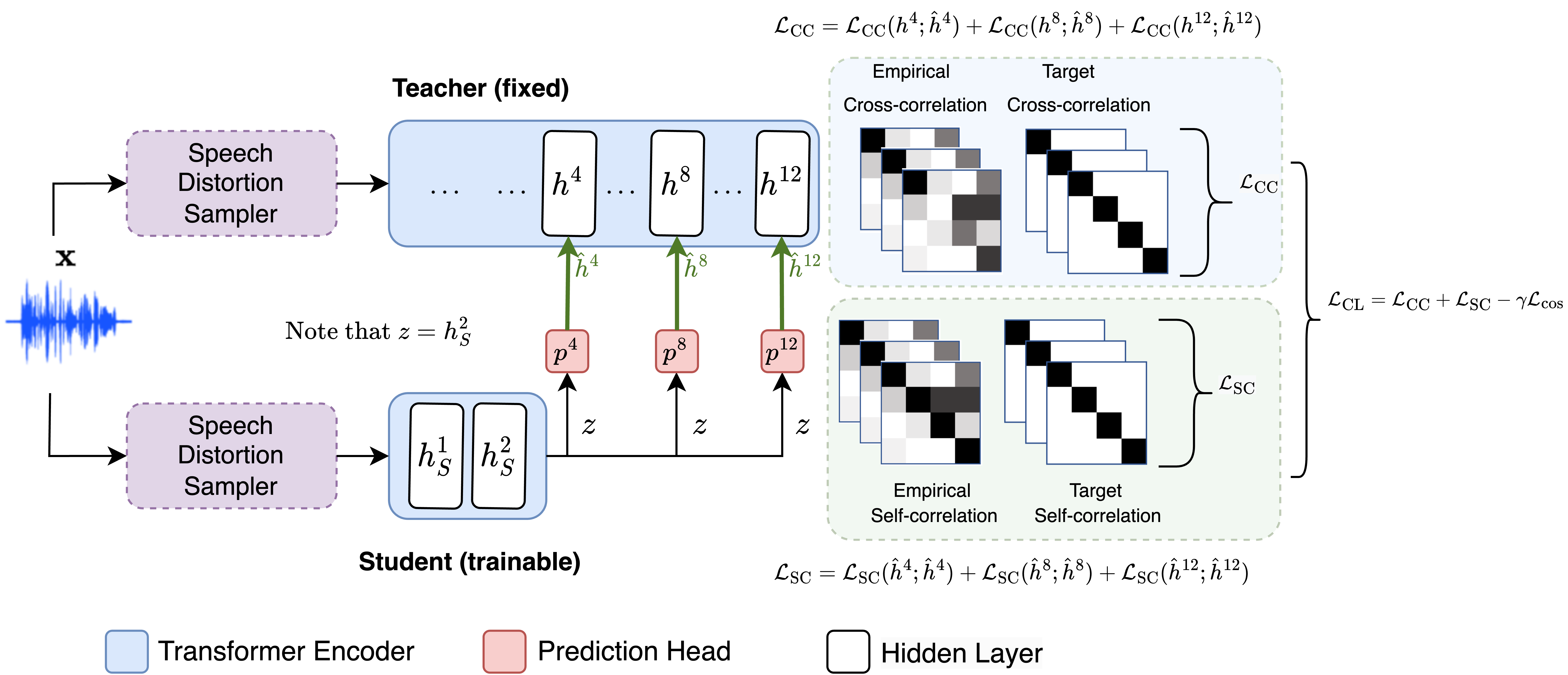}
  \caption{Illustration of our distillation framework. The Speech Distortion classifier follows the setup described in Section \ref{subsec:pretrain}. Our method minimizes the self- and cross-correlation matrix between the embeddings of the teacher and the student.}
  \label{fig:distilation_overview}
\end{figure*}

\subsection{Barlow Twins}
Barlow Twins \cite{barlowtwins} (BT) is a self-supervised learning technique proposed in CV that aims at learning augmentation invariant representation while maximizing the information learned on the feature dimension in the representations of the model.  BT feeds two distorted views of an image to the same neural network architecture. The training objective maximizes the diagonal elements of the cross-correlation matrix between the representations of the two input views as,

\begin{equation}
\label{eq:barlow_twins_original}
\mathcal{L}_{\text{BT}} = \sum_{i=1}^{D'}\left(1-c_{i i}\right)^2 +\lambda \sum_{i=1}^{D'} \sum_{j=1,j \neq i}^{D'} c_{i j}^2,
\end{equation}
where $\mathcal{C}= \{c_{ij}, \forall_{i,j} \in 1,2,..., D^\prime\} $ is the cross-correlation matrix computed between the outputs of the network along the batch dimension. In \cite{barlowtwins}, the cross-correlation computation assumes that the features have been flattened and projected towards a high dimensional vector of dimension $D' \gg D$. Namely, an element $c_{ij}$ is computed as,

\begin{equation}
\label{eq:c_i_j_bt}
c_{i j} = \frac{\sum\limits_{b=1}^{B} y_{b, i}^{V1} y_{b, j}^{V2}}{\sqrt{\sum\limits_{b=1}^B\left(y_{b, i}^{V1}\right)^2} \sqrt{\sum\limits_{b=1}^{B}\left(y_{b, j}^{V2}\right)^2}},
\end{equation}
where $\mathbf{y}^{V1}$, $\mathbf{y}^{V2} \in \mathbb{R}^{B \times D'}$, with $B$ equals to the batch size and $D'$ the high-dimensional feature dimension, represents a batch of embeddings from the same input signal that has been distorted/augmented with the distortion ${V1}$ and ${V2}$ respectively. The value of $D'$ is defined by the size of the projector layer, hence $\mathcal{C} \in \mathbb{R}^{D' \times D'}$.

\section{Proposed method}
\subsection{Correlation loss for SSL distillation}

 While the conventional distillation loss of DistilHuBERT between a clean representation on the teacher model and a distorted representation on the student may enforce noise invariant representations, it doesn't always prevent the student from learning distorted representations if the teacher isn't distortion invariant. Furthermore, having a method that can encourage maximally informative representations on DistilHuBERT can benefit DistilHuBERT's performance in both noisy scenarios and clean downstream taks. Recall that in SUPERB \cite{superb}, when an SSL model is used for a downstream task, the SSL model is frozen, and only a weight vector is trained to learn a linear combination of the SSL hidden layers. Therefore, for a HuBERT model, there is a 12-dimensional weight vector to learn. In \cite{superb,layerwiselivescu}, it is shown that different layers are better at localizing different types of information, meaning that different layers are more useful for specific downstream tasks than others. Nonetheless, DistilHuBERT only uses the last hidden layer for downstream tasks, implying that apart from having noise invariant representation, having features that maximize their information across the feature dimension could benefit noise-robustness and downstream generalization, this desirability is achieved by the BT objective. 
 
 Using BT loss directly for knowledge distillation is impractical. First, Barlow Twins was not designed for sequential data. Previous work on speech \cite{bt_speaker_rec,bt_er} either used Barlow Twins by averaging the time dimension into a single vector or via concatenation of the cropped sequential data, subsequently flattening the matrix and passing the representation to a high-dimensional projector layer. Time dimension averaging is not beneficial for SSL downstream performance as observed in our preliminary investigations. A second challenge of this work is that there is no siamese architecture involved but rather different architecture topologies for the teacher and the student. This mismatch poses an interesting challenge. Consider the case where the representations of the teacher model are distorted. In this case, because the teacher is frozen, the representations of the teacher cannot change, and hence the off-diagonal elements of the cross-correlation matrix between the teacher and the student are limited by the level of disentanglement of the already trained teacher model. Additionally, there is a case where minimizing the off-diagonal elements of the cross-correlation term in the original BT objective does not guarantee distortion-invariance. For the case where the teacher outputs distortion-invariant representations, even if the student outputs representations containing distorted information, the cross-correlation off-diagonal elements may still remain low. In this case, the student model will have to rely merely on the diagonal elements in order to output distortion-invariant representations. This problem can be avoided by adding a self-correlation term to the student model.

Hence, this paper proposes to improve noise robustness and downstream generalization by exploiting cross and self-correlation matrices over the teacher-student distillation framework. Namely, let $\hat{H} \in \mathbb{R}^{B \times P \times T \times D}$ be the predicted hidden representations of the student model. $P$ denoting the number of prediction heads. Similarly, let $H \in \mathbb{R}^{B \times P \times T \times D}$ represent the hidden representations of the teacher.
Assuming the representations have been mean-normalized over the batch dimension, we compute the cross-correlation matrix as,

\begin{equation}
\label{eq:cc_loss}
\mathcal{C}_{\text{cc}} = \frac{1}{B}\sum_{b=1}^B \hat{H}_{b}^{\prime} H_{b}^{\prime},
\end{equation}
with $\hat{H}^{\prime}\in \mathbb{R}^{B \times P \times T \times D \times 1}$, the unsqueezed and mean-normalized student predicted hidden representations, $H^{\prime} \in \mathbb{R}^{B \times P \times T \times 1 \times D}$ the teacher one, and $\mathcal{C}_{\text{cc}} \in  \mathbb{R}^{P \times T \times D \times D}$. The self-correlation term $\mathcal{C}_{\text{sc}}$, follows the same logic as Eq. (\ref{eq:cc_loss})
by calculating the matrix multiplication of $\hat{H}$ by itself. Namely,

\begin{equation}
\label{eq:sc_loss}
\mathcal{C}_{\text{sc}} = \frac{1}{B}\sum_{b=1}^B \hat{H}_{b} \hat{H}_{b},
\end{equation}

In Eq. \eqref{eq:sc_loss}, $\hat{H}$ is as well unsqueezed and mean normalized so that $\mathcal{C}_{\text{sc}} \in  \mathbb{R}^{P \times T \times D \times D}$.

Combining both terms, the correlation objective $\mathcal{L}_{\text{CL}}$ is,

\begin{align}
\label{eq:CL}
\mathcal{L}_{\text{CL}} &= \mathcal{L}_{\text{CC}} + \mathcal{L}_\text{{SC}} - \gamma  \mathcal{L}_{\text{cos}}\\
\mathcal{L}_{\text{CC}} &= \sum_i\left(1-\mathcal{C}_{\text{cc}_{i i}}\right)^2 +\lambda_{cc} \sum_i \sum_{j \neq i} \mathcal{C}^{2}_{\text{cc}_{i j}}\\
\mathcal{L}_\text{{SC}} &= \lambda_{sc} \sum_i \sum_{j \neq i} \mathcal{C}^{2}_{\text{sc}_{i j}}
\end{align}

Note that this implementation does not need a flatten operation and a projector layer as done in the original BT objective \cite{barlowtwins} or in related work \cite{ng2023hubert,delores,bt_speaker_rec}. Hence, no additional parameters are required to perform distillation. An illustration of the proposed method can be seen in Fig. \ref{fig:distilation_overview}.

\section{Experiments}\label{sec:experiments}

\begin{table*}[t!]
\setlength\tabcolsep{1.4pt}
\renewcommand{\arraystretch}{0.8}
\centering
\caption{Accuracy (Acc\% ↑) results on IC and KS and Word Error Rate (WER\% ↓) results for ASR task on the SUPERB Benchmark for the previous approach ($\mathcal{L}_{\text{KD}}$) versus our proposed method ($\mathcal{L}_{\text{CL}}$) on HuBERT and HuBERT+ teachers. Clean refers to the unmodified SUPERB test set, while noisy refers to the distorted set with CHiMe3 OOD noise. $\mathcal{L}_{\text{CL}}$ uses $\lambda_{cc}=5\mathrm{e}{-5}$ and $\lambda_{sc}=5\mathrm{e}{-6}$.} 
\label{tab:setup1_vs_setup2}
\begin{tabular}{llcccc|ccc|ccc|ccc}
    \toprule
    
     & \multicolumn{1}{l}{} & \multicolumn{1}{l}{} & \multicolumn{1}{l}{} & \multicolumn{1}{c}{} & \multicolumn{1}{c}{}  &\multicolumn{3}{|c}{\textbf{IC} (Acc$\%\uparrow$)} & \multicolumn{3}{|c}{\textbf{KS} (Acc$\%\uparrow$)} & \multicolumn{3}{|c}{\textbf{ASR} (WER$\%\downarrow$)}  \\

 \multicolumn{1}{l}{Block} & \multicolumn{1}{l}{Upstream} & \multicolumn{1}{l}{\#params} & Pretrain & \multicolumn{1}{l}{Distorted} & \multicolumn{1}{l}{Loss} & \multicolumn{3}{|c}{} & \multicolumn{3}{|c}{} & \multicolumn{2}{|c} {} & \multicolumn{1}{c}{}  \\

   & & \multicolumn{1}{c}{(M)} & \multicolumn{1}{c}{Dataset} & \multicolumn{1}{c}{Input}  &\multicolumn{1}{c}{Type}   &\multicolumn{1}{|c}{clean}  &  \multicolumn{1}{c}{noisy} & \multicolumn{1}{c}{diff$\downarrow$}  & \multicolumn{1}{|c}{clean} &  \multicolumn{1}{c}{noisy} & \multicolumn{1}{c}{diff$\downarrow$}  & \multicolumn{1}{|c}{clean}  &  \multicolumn{1}{c}{noisy} & \multicolumn{1}{c}{diff$\downarrow$}  \\
  \midrule
  & \multicolumn{2}{l}{Baselines}\\
  \midrule
   \multirow{4}{*}{1} & HuBERT Base & 95 & LS960 & None & - & 98.34 & 90.96 &  7.38 & 96.30 &93.74  & 2.56 & 6.42 & 8.58 & 2.16\\
    & HuBERT+ Base \cite{kuanpocontinualtraining} & 95 & LS960 & None & - &  98.37 & 97.02 & 1.35 & 96.17 & 94.94 &1.23 & 6.97 & 9.41 &2.44  \\
   & \textbf{DistilHuBERT} & 23& \textbf{LS100} & None   & $\mathcal{L}_{\text{KD}}$   & 91.20&	66.71&24.49 &  95.12&	90.29&4.83	& 15.83&	29.65 & 13.82\\
    \midrule
  & \multicolumn{4}{l}{Noise Robust Distilled Methods}\\
    \midrule
  \multirow{2}{*}{2} & DistilHuBERT & 23& LS100   & Student      &  $\mathcal{L}_{\text{KD}}$   &  92.43	& 86.59 &5.84  & 95.22 & 93.26  &1.96  & 16.37	& 19.84	& 3.47 \\
   & DistilHuBERT &23 &LS100  &  Student          &  $\mathcal{L}_{\text{CL}}$                & 95.73	   & 91.25 &  4.48 & \textbf{95.72}	& 94.16 & 1.56 & 15.62 & 18.78	&	3.16 \\
 \midrule
 \multirow{2}{*}{3}  & DistilHuBERT &23 & LS100  &  Both         &  $\mathcal{L}_{\text{KD}}$   & 93.03	& 89.29 & 3.74 &  95.41	& 93.15 & 2.26 & 16.98 & 20.47  & 3.49 \\
   
   & DistilHuBERT &23 &  LS100  &  Both         &  $\mathcal{L}_{\text{CL}}$   & \textbf{96.61} & \textbf{94.06} & \textbf{2.55} &  95.59 & \textbf{94.38} & \textbf{1.21} & 15.35 & 18.18 & 2.83  \\
\midrule
 \multirow{2}{*}{4}  & DistilHuBERT+ &23 & LS100  &  Student                      &  $\mathcal{L}_{\text{KD}}$   &  93.92 & 87.13 & 6.81 & 95.55	 & 91.97 &  3.58 & 16.27  & 19.59 & 3.32  \\

   & DistilHuBERT+ &23 &  LS100  &  Student                      &  $\mathcal{L}_{\text{CL}}$  & 95.20 &90.99 & 4.21  & 95.62	 & 93.61 &  2.01&   \textbf{15.31} & 17.99 &	2.68 \\
\midrule
  \multirow{2}{*}{5} & DistilHuBERT+ &23 & LS100  &  Both                      &  $\mathcal{L}_{\text{KD}}$   &  95.48  & 90.34  & 5.14 & 95.60	& 93.81 & 1.79 & 16.77	& 20.74	& 3.97  \\
    
   & DistilHuBERT+ &23 & LS100  &  Both                     &  $\mathcal{L}_{\text{CL}}$                & 96.07 & 93.04 & 3.03 & 95.68 & 93.83 & 1.85 & 15.32 & \textbf{17.77} & 2.45  \\

\bottomrule
\end{tabular}
\end{table*}

\begin{table}[t!]
\setlength\tabcolsep{2pt}
\renewcommand{\arraystretch}{0.8}
\centering
\caption{Performance of DistilHuBERT with the heuristic method for $\lambda_{cc}$ and $\lambda_{sc}$. The first row refers to experiments using $\mathcal{L}_{\text{KD}}$ loss. The second row is the method without automatic weighting. The third row refers to the heuristic method proposed. Both teacher and student receive speech distortions.}
\label{tab:automatic_lambda}
\begin{tabular}{l|cc|cc|cc}
    \toprule
    
       \multicolumn{1}{l}{Off-diagonal}   & \multicolumn{2}{|c}{\textbf{IC} (Acc$\%\uparrow$)} & \multicolumn{2}{|c}{\textbf{KS} (Acc$\%\uparrow$)} & \multicolumn{2}{|c}{\textbf{ASR} (WER$\%\downarrow$)}  \\

   \multicolumn{1}{l}{Weighting} & \multicolumn{2}{|c}{} & \multicolumn{2}{|c}{} & \multicolumn{2}{|c} {} \\

     \multicolumn{1}{l}{} & \multicolumn{1}{|c}{clean}  &  \multicolumn{1}{c}{noisy}  & \multicolumn{1}{|c}{clean} &  \multicolumn{1}{c}{noisy}  & \multicolumn{1}{|c}{clean}  &  \multicolumn{1}{c}{noisy} \\
 \midrule
       None ($\mathcal{L}_{\text{KD}}$) & 93.03	& 89.29 &  95.51	& 93.15  & 16.98 & 20.47   \\
   
   Fixed & 96.61 & 94.06 & 95.59 & 94.38 & 15.35 & 18.18 \\
     Heuristic &\textbf{96.63}	&\textbf{94.12}	&\textbf{95.85}	&\textbf{94.96}	&\textbf{14.89}	&\textbf{17.54}	\\
   
   \bottomrule
\end{tabular}
\end{table}

    

\subsection{Pre-training Data Description}\label{subsec:pretrain}
This work follows \cite{RobustDistilHuBERT} data configuration for direct performance comparison. The training data for knowledge distillation is the 100-hour clean Librispeech subset \cite{librispeech}. Noises are added to the speech data following the configuration of \cite{RobustDistilHuBERT}. Namely, two setups are considered: \textbf{Setup 1} regards pre-training by adding noise distortion only to the input of the student model, while \textbf{Setup 2} adds different types of noises to the teacher and student input. In all the setups mentioned, the clean speech is distorted by a combination of additive and non-additive distortions. For additive distortions, noise from datasets Musan \cite{musan}, WHAM! \cite{wham} as well as Gaussian perturbation are added to the speech training data at a signal-to-noise ratio (SNR) ranging in $[10, 20)$ dB. For non-additive distortions, reverberation, pitch shift and band rejection are applied to the speech training data. 

\subsection{Teacher-Student Pre-training}

This work adopts HuBERT (HB) base model as the teacher model and HUBERT+ (HB+) \cite{kuanpocontinualtraining} to assess the generalizability of the proposal to different teacher models. The network architecture of the student model is the same as DistilHuBERT, regardless of the types of the teacher model. When distilling HB or HB+, the student parameters are initialized from the CNN layers and the first two transformer layers of the teacher model. For knowledge distillation, each model is trained for 200k steps, and the selected checkpoint is the model step with the lowest pre-training loss on the dev-clean set of LibriSpeech.

\subsection{Downstream Training and test-set Evaluation}

During downstream training, the student model parameters are frozen, and only the representations of the last hidden layer are used as the input of the downstream models. Other hyperparameters and configurations, such as batch size, training steps and downstream model architectures follow the same configuration of SUPERB Benchmark \cite{RobustDistilHuBERT,superb}. For downstream performance under clean and noisy settings, results on IC, KS and ASR tasks are reported. We report only out-of-distribution (OOD) noise to assess generalization to unseen noise. The OOD noise consists of a noise perturbation of the original clean speech with CHiMe3 noise \cite{chime3}. CHiMe3 noise is added to the testing data following the approach in \cite{huang2023ensemble}. Namely, the background noises of CHiMe3 dataset are used as additive noise to the original clean speech. Before performing the speech perturbation, segmentation on the background noise audio is done to avoid adding portions with silences only. This method is denoted as ``noisy" in Table \ref{tab:setup1_vs_setup2} and Table \ref{tab:automatic_lambda}.

\section{experimental Results}\label{sec:results}
\subsection{Different teacher models}\label{sec:diffteachers}

This section aims to analyze the performance of the proposed approach $\mathcal{L}_{\text{CL}}$ versus the previous $\mathcal{L}_{\text{KD}}$ under different teacher models to compare the case of having a non-noise robust teacher model (HuBERT) and a noise robust one (HuBERT+).

Table \ref{tab:setup1_vs_setup2} shows the performance of the proposed method under different teacher topologies and noise perturbations. In Table \ref{tab:setup1_vs_setup2}, all the experiments for the proposed method use a weight of $\lambda_{cc}=5\mathrm{e}{-5}$ and $\lambda_{sc}=5\mathrm{e}{-6}$ for the off-diagonal elements of the cross and self-correlation matrix. The values have been chosen based on preliminary hyperparameter optimization for the ASR task under the setup where the teacher receives clean speech and the student receives distorted speech. We found out that this setup provided the best dev-set performance on Librispeech.

Block 1 in Table \ref{tab:setup1_vs_setup2} presents some baseline results. From this Block, it can be seen the considerable performance degradation of DistilHuBERT when the data is corrupted with chime noise. Particularly, there is an absolute decrease in performance of 24.49\% and 4.83\% on IC and KS tasks compared to a decrease of  7.38\% and 2.56\%, respectively on the HuBERT Base teacher model.

Blocks 2 and 3 in Table \ref{tab:setup1_vs_setup2} compare the proposed method $\mathcal{L}_{\text{CL}}$ with the previous approach $\mathcal{L}_{\text{KD}}$ under the setup where only the student receives noise distortion and the setup where both teacher and student receives distortions for a HuBERT teacher model. From Block 2, it can be seen that the proposed approach improves the previous method by 3.3\% absolute improvement on the clean set of the IC task and 4.66\% on the chime noise perturbation scenario. Consistent improvement is also observed across KS and ASR tasks. Furthermore, our proposed method also reduces the gap between clean and noise performance from  5.84\% absolute performance degradation to 4.48\% on IC. The improvement of the proposed method for HuBERT distillation when both teacher and student receive distortion (Block 3 in Table \ref{tab:setup1_vs_setup2}) is also significant with 3.58\% and 4.75\% absolute improvement on the clean and chime noise scenario on IC. Again, the proposed method is also able to reduce the performance degradation gap between clean and OOD noise scenarios. Similar trends are observed when the distillation is done on a noise roust teacher (HuBERT+) where our method consistently improves over the previous approach. Finally, it is interesting to note that our proposed method achieves overall the best clean and noise performance when doing Distillation on the original HuBERT model rather than in the noise robust HuBERT+ model. This finding suggests that the proposed approach is agnostic of the teacher model and that noise generalization can be achieved even if the teacher is not noise robust. Similarly it suggest that our method benefits further from a non noise robust teacher. More analyses are presented in Section \ref{sec:analysis_representations}.

\begin{figure}[!th]
  \centering
  \includegraphics[width=0.92\linewidth]{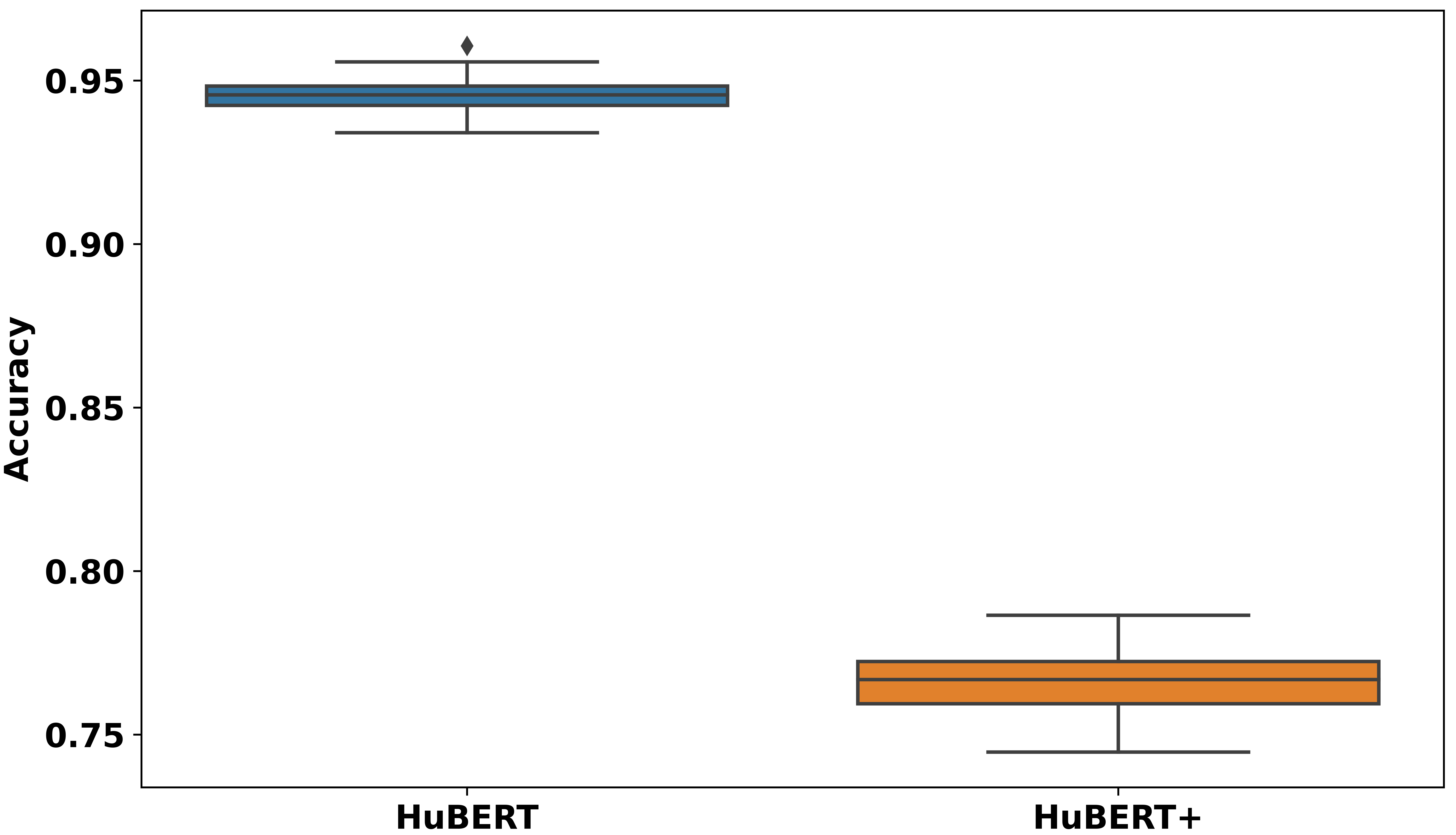}
  \caption{Noise classification accuracy for HuBERT (blue) and HuBERT+ (orange).}
  \label{fig:noise_acc_hub_hubplus}
\end{figure}

\begin{figure}[!th]
  \centering
  
  \begin{subfigure}[b]{0.82\linewidth}
    \includegraphics[width=\linewidth]{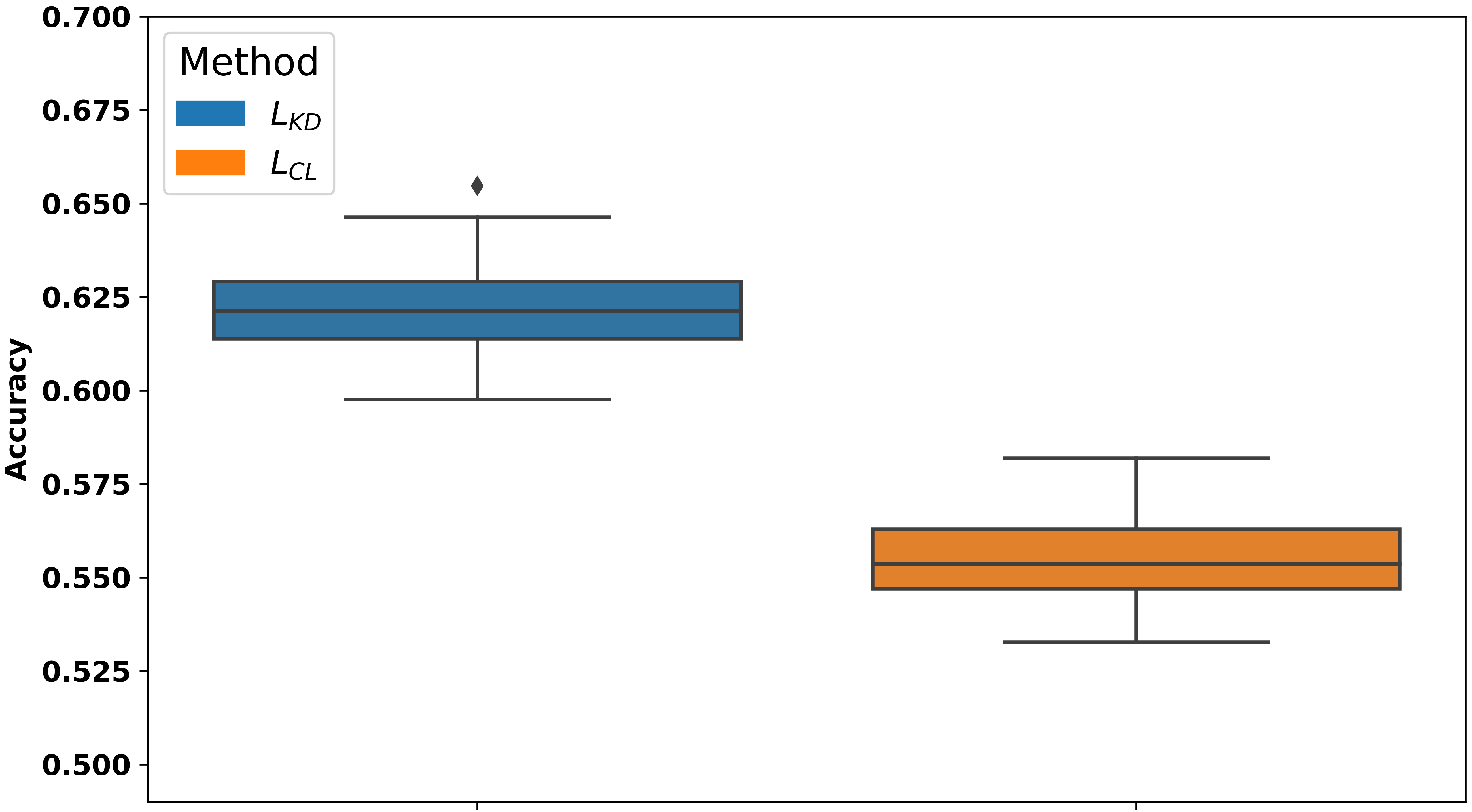}
        \caption{Noise accuracy for HuBERT distillation.}

    \label{fig:hubertdistil}
  \end{subfigure}
  \hfill 
  \begin{subfigure}[b]{0.82\linewidth}
    \includegraphics[width=\linewidth]{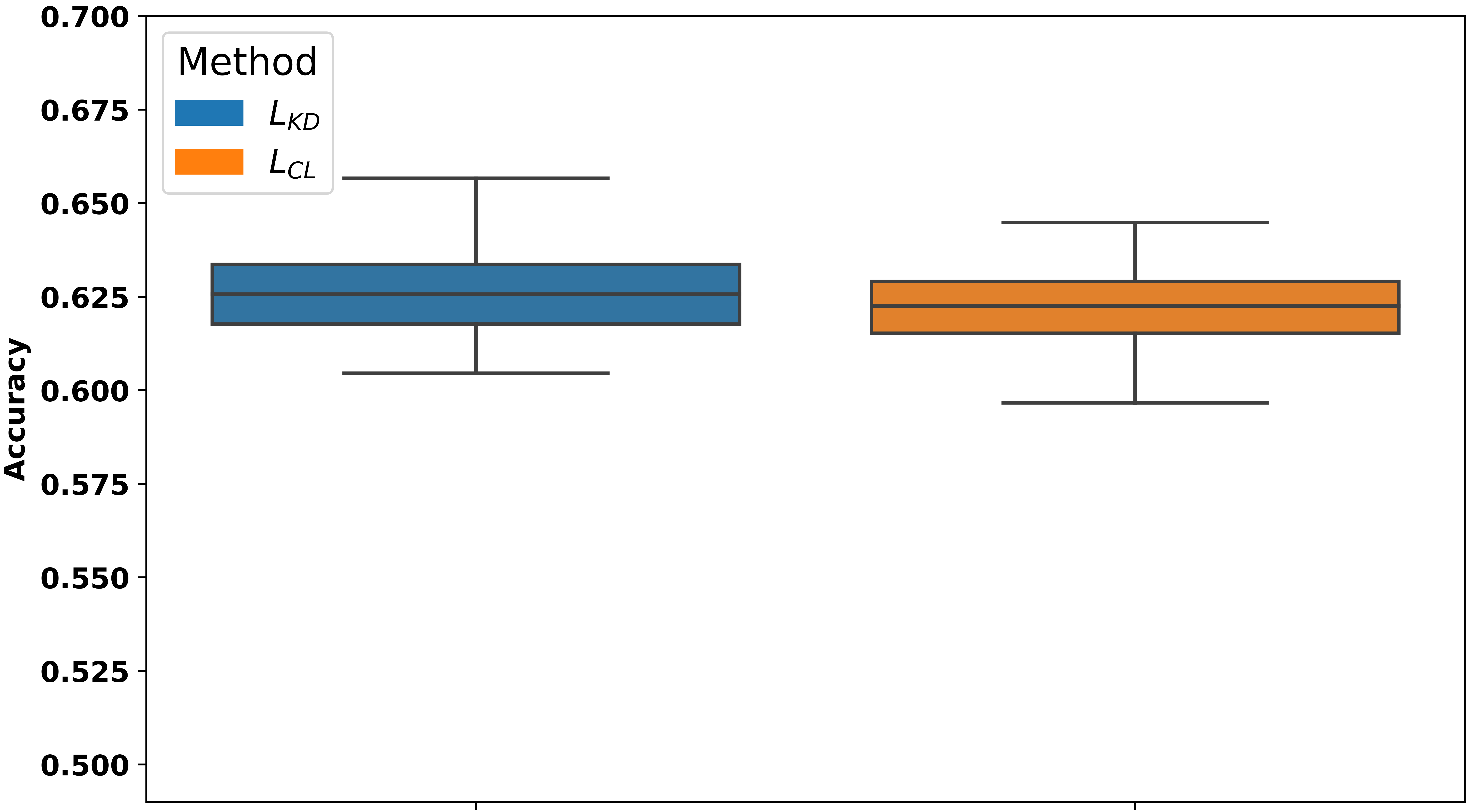}
    \caption{Noise accuracy for HuBERT+ distillation.}
    \label{fig:hubertplusdistil}
  \end{subfigure}
  
  \caption{Noise classification accuracy for distillation of HuBERT (top) and HuBERT+ (bottom) with $\mathcal{L}_{\text{KD}}$ (blue) and $\mathcal{L}_{\text{CL}}$ (orange). Models correspond to the models in Block 3 (top plot) and Block 5 (bottom plot) in Table \ref{tab:setup1_vs_setup2}.}
  \label{fig:noise_acc_distillation}
\end{figure}

\begin{table}[htbp]
\renewcommand{\arraystretch}{0.6}
\centering
\caption{Layer analysis on clean Librispeech 100 hour subset for transformer layers T4, T8, and T12 for HuBERT and HuBERT+. The same protocol as in \cite{layerwiselivescu,livescu_all_models} is followed.}
\label{tab:layerwiseanalysis}
\begin{tabular}{
    l
    l
    S[table-format=1.3]
    S[table-format=1.3]
    S[table-format=1.3]
}
\toprule
Model& Metric & {T4} & {T8} & {T12} \\
\midrule
\multirow{3}{*}{HuBERT} & CCA-mel & 0.630 & 0.620 & 0.660 \\
& MI-phone & 0.896 & 0.897 & 0.885 \\
\midrule
\multirow{3}{*}{HuBERT+} & CCA-mel & 0.640 & 0.600 & 0.610 \\
& MI-phone & 0.881 & 0.913 & 0.895 \\
\bottomrule
\end{tabular}
\end{table}

\subsection{Automatic cross and self-correlation weight term} 

While doing hyper-parameter optimization for $\mathcal{L}_{\text{CL}}$, we find out that there is a trade-off between clean and noise robustness performance depending on the manually chosen $\lambda_{cc}$ and $\lambda_{sc}$ terms for the cross and self-correlation computation. We found out that giving more importance to the self-correlation term improves clean and noise robustness when the student is receiving distorted representations. For this reason, in this section, we aim to assess if using an automatic mechanism to set the coefficients for $\lambda_{cc}$ and $\lambda_{sc}$ helps with noise generalization. We call this method the ``heuristic" approach. This heuristic approach consists of a simple probe to validate the hypothesis that using different $\lambda$'s terms during pre-training can make the model better at handling noise. This differs from the previous $\mathcal{L}_{\text{CL}}$ computation where the $\lambda_{cc}$ and $\lambda_{sc}$ coefficients are fixed during pre-trainig. The method follows a simple heuristic as follows: At pre-training, an SNR from [10,20) is chosen either for the teacher, for the student, or both. Hence, considering this SNR value is known, we compute a $\lambda$ parameter that changes linearly from an interval of [5$\mathrm{e}{-5}$,5$\mathrm{e}{-7}$). Particularly, given a SNR value $ \mathbf{s} \in [10,20)$,

\begin{equation}\label{eq:heuristic}
    \lambda = 5\times \mathrm{10}^{-5}  (9.9  \mathbf{s} - 98),
\end{equation}

with $\lambda \in [5\times \mathrm{10}^{-5},5\times \mathrm{10}^{-7})$. Here, both $\lambda_{cc}$ and $\lambda_{sc}$ are weighted under this heuristic where $\lambda_{sc}$ is weighted based on the SNR of the student input and $\lambda_{cc}$ based on the SNR of the teacher input. From Eq. \eqref{eq:heuristic}, it can be observed that this method will give more importance to the $\lambda$ parameter whenever the SNR is low, meaning whenever the input has high distortion, while lower importance is given when the signal is clean.

Table \ref{tab:automatic_lambda} shows the results of such an approach for the case where both teacher and student receive distortion. From the results in Table 
 \ref{tab:automatic_lambda}, it can be observed that this simple technique can improve clean and noise generalization considerably than with the manually fixed $\lambda$ setup. Future work will explore this direction in more detail.

\subsection{Understanding the effect of the teacher model on distillation performance}\label{sec:analysis_representations}

Section \ref{sec:diffteachers} shows that our proposed method consistently outperforms previous noise robust distillation method \cite{RobustDistilHuBERT} on each of the block comparisons shown in Table \ref{tab:setup1_vs_setup2}. Nonetheless, the best model is achieved when our proposed method distills from the original HuBERT model rather than HuBERT+. In order to shed light on this finding, an analysis of the noise invariancy of the representations is done as follows.

First, we iterate over the SSL embeddings of the train-100 Librispeech subset and create 4 versions of the embeddings, each of them perturbed with one kind of noise only. Particularly we create 4 versions by perturbing the whole subset with chime background noise, FSD50k noise \cite{FSD50K}, reverberation and gaussian noise, respectively. We then meanpool the representations by utterance and construct a noise classifier using a Random Forest with a bootstrap size of 100. The results of the noise classification for HuBERT and HuBERT+ model can be seen in Fig. \ref{fig:noise_acc_hub_hubplus}, showing that HuBERT has almost 95\% of noise classification accuracy while HuBERT+ has around 76\%. Fig. \ref{fig:noise_acc_hub_hubplus} results show that HuBERT is noise variant while HuBERT+ is more noise invariant as the features carry less noise information. Hence it achieves lower noise classification accuracy. On the other hand, Fig. \ref{fig:noise_acc_distillation} shows the noise classification accuracy of the distilled models using 
$\mathcal{L}_{\text{KD}}$ (blue) and the proposed method $\mathcal{L}_{\text{CL}}$ (orange) when the teacher model is HuBERT (top) and when the teacher model is HuBERT+ (bottom). These results suggest some interesting findings. First of all, it can be noticed that using the standard $\mathcal{L}_{\text{KD}}$ method reaches almost the same level of noise invariancy for both teacher models (62.5\% accuracy). On the other hand, our method benefits the most for the case where the teacher model is not noise robust. This claim is supported by the lower noise accuracy reached for the model that distills from HuBERT rather than from HuBERT+, explaining why the better performance of the proposed approach in Block 3 vs the proposed approach in Block 5 in Table \ref{tab:setup1_vs_setup2}. The intuition is that distilling from a noise variant teacher makes the off-diagonal minimization of the cross and self-correlation matrix more relevant, while if the teacher model is noise robust, focusing more on the minimization of the diagonal elements of the cross-correlation matrix is enough. These results help in understanding the capability of the proposed approach to be agnostic of the teacher model no matter the teacher noise robustness. We conclude that the proposed approach benefits more from a noise variant teacher model while it can still outperform the previous approach in the scenario where the teacher model is noise invariant. Finally, to validate this last claim and to isolate the noise variance/invariance as the reason for different performances on distillation, we also proceed to do a layer-wise analysis of both teacher models following the exact same configurations of \cite{layerwiselivescu,livescu_all_models}. Table \ref{tab:layerwiseanalysis} shows such results where it can be observed that both teacher models have negligible differences in CCA-mel (similarity of the representations with mel-spectrogram features) and MI-phone (the ability of the representations to encode phonetic information) modeling capabilities.

\section{Conclusions}

This paper has proposed a new correlation-based objective for improved clean and noise-robust distilled speech foundation models. The proposed method maximizes the diagonal elements of the cross-correlation matrix between the representations of the teacher and student while it also minimizes the off-diagonal elements of the cross and self-correlation matrix. The proposed method significantly improves clean and noisy conditions for different teacher models while also reducing the gap of performance degradation between clean and unseen noise. Besides, this new distillation framework has been shown to be noise robust even in the case where the teacher model is not robust against noise allowing our method to be agnostic on the teacher model characteristics. Finally, this paper has studied the possibility of automatically tuning the interpolation weights of the off-diagonal terms on the cross and self-correlation matrix, which establishes a path for further work in this direction. Future work will analyze methods to speed up the computation of the correlation terms and explore the use of sequence-level compression techniques with this correlation-based objective. 

\bibliographystyle{IEEEtran}
\bibliography{mybib}

\end{document}